\documentclass{JHEP3}   
\usepackage{epsfig}

\newcommand{\eref}[1]{(\ref{#1})}

\newcommand{\fref}[1]{Figure~\ref{#1}}
\newcommand{\cref}[1]{Chapter~\ref{#1}}
\newcommand{\beq}{\begin{equation}}
\newcommand{\eeq}{\end{equation}}
\newcommand{\ba}{\begin{array}}
\newcommand{\ea}{\end{array}}
\newcommand{\bcenter}{\begin{center}}
\newcommand{\ecenter}{\end{center}}

\def\IB{\relax\hbox{$\inbar\kern-.3em{\rm B}$}}
\def\IC{\relax\hbox{$\inbar\kern-.3em{\rm C}$}}
\def\ID{\relax\hbox{$\inbar\kern-.3em{\rm D}$}}
\def\IE{\relax\hbox{$\inbar\kern-.3em{\rm E}$}}
\def\IF{\relax\hbox{$\inbar\kern-.3em{\rm F}$}}
\def\IG{\relax\hbox{$\inbar\kern-.3em{\rm G}$}}
\def\IGa{\relax\hbox{${\rm I}\kern-.18em\Gamma$}}
\def\IH{\relax{\rm I\kern-.18em H}}
\def\IK{\relax{\rm I\kern-.18em K}}
\def\IL{\relax{\rm I\kern-.18em L}}
\def\IP{\relax{\rm I\kern-.18em P}}
\def\IR{\relax{\rm I\kern-.18em R}}
\def\IZ{\relax\ifmmode\mathchoice
{\hbox{\cmss Z\kern-.4em Z}}{\hbox{\cmss Z\kern-.4em Z}}
{\lower.9pt\hbox{\cmsss Z\kern-.4em Z}}
{\lower1.2pt\hbox{\cmsss Z\kern-.4em Z}}\else{\cmss Z\kern-.4em Z}\fi}
\def\II{\relax{\rm I\kern-.18em I}}

\def\sCC{{\kern 0.27em\vrule height1.45ex width0.03em depth0em
          \kern-0.30em\rm C}}
\def\C{{\mathchoice
  {\sCC}
  {\sCC}
  {\kern 0.225em \vrule height1.05ex width0.025em depth0em \kern-0.25em \rm C}
  {\kern 0.180em \vrule height0.78ex width0.02em depth0em \kern-0.2em \rm C}
        }}
\def\sHH{{\rm I\kern-.16em{}H}}
\def\H{{\mathchoice
  {\sHH}
  {\sHH}
  {\rm I\kern-.13em{}H}
  {\rm I\kern-.13em{}H} }}
\def\sNN{{\rm I\kern-.16em{}N}}
\def\N{{\mathchoice
  {\sNN}
  {\sNN}
  {\rm I\kern-.12em{}N}
  {\rm I\kern-.10em{}N} }}
\def\sPP{{\rm I\kern-.16em{}P}}
\def\P{{\mathchoice
  {\sPP}
  {\sPP}
  {\rm I\kern-.12em{}P}
  {\rm I\kern-.10em{}P} }}
\def\sQQ{{\kern 0.27em \vrule height1.45ex width0.03em depth0em
          \kern-0.30em \rm Q}}
\def\Q{{\mathchoice
        {\sQQ}
        {\sQQ}
  {\kern 0.225em \vrule height1.05ex width0.025em depth0em \kern-0.25em \rm Q}
  {\kern 0.180em \vrule height0.78ex width0.020em depth0em \kern-0.20em \rm Q}
        }}
\def\sRR{{\rm I\kern-0.16em{}R}}
\def\R{{\mathchoice
  {\sRR}
  {\sRR}
  {\rm I\kern-0.12em{}R}
  {\rm I\kern-0.10em{}R} }}
\def\sZZ{{\rm Z\kern-0.32em{}Z}}
\def\Z{{\mathchoice
  {\sZZ}
  {\sZZ} 
  {\rm Z\kern-0.3em{}Z}     
  {\rm Z\kern-0.25em{}Z} }}  
\def\ZZZ{{\rm Z\kern-0.24em{}Z}}
\def\sII{{\rm I\kern-0.16em{}I}}
\def\I{{\mathchoice
  {\sII}
  {\sII}
  {\rm I\kern-0.12em{}I}
  {\rm I\kern-0.10em{}I} }}

\def\vol{{\rm vol}}

\def\inbar{\,\vrule height1.5ex width.4pt depth0pt}
\font\cmss=cmss10 \font\cmsss=cmss10 at 7pt

\def\smiley{\hbox{\large$\bigcirc$\hspace{-0.80em}\raise.2ex
\hbox{$\cdot\cdot$}\kern-.61em\lower.2ex\hbox{\scriptsize$\smile$}}\ }
\def\frowny{\hbox{\large$\bigcirc$\hspace{-0.80em}\raise.2ex
\hbox{$\cdot\cdot$}\kern-.635em\lower.2ex\hbox{\scriptsize$\frown$}}\ }

\def\I{{\rlap{1} \hskip 1.6pt \hbox{1}}}

\makeatletter
\let\hangafter\@hangfrom
\makeatother

\newcommand{\be}{\begin{equation}}
\newcommand{\ee}{\end{equation}}
\newcommand{\bea}{\begin{eqnarray}}
\newcommand{\eea}{\end{eqnarray}}
\newcommand{\bean}{\begin{eqnarray*}}
\newcommand{\eean}{\end{eqnarray*}}
\newcommand{\nn}{\nonumber}

\newcommand{\mB}{\mathcal{B}}

\preprint{MIT-CTP-3570, CERN-PH-TH/2004-235, HUTP-04/A0050}

\title{An Infinite Family of Superconformal Quiver Gauge Theories with Sasaki-Einstein Duals}

\author{Sergio Benvenuti$^1$, Sebasti\'an Franco$^2$,  Amihay Hanany$^2$, Dario 
Martelli$^3$, James Sparks$^4$\\

\vspace{0.3 cm}

{1. Scuola Normale Superiore, Pisa,
                   and INFN, Sezione di Pisa, Italy.}\\
~\\
{2. Center for Theoretical Physics,
                   Massachusetts Institute of Technology,\\
                   Cambridge, MA 02139, USA.}\\
~\\
{3. Department of Physics, CERN Theory Division,
                   1211 Geneva 23, Switzerland.}\\
~\\
{4. Department of Mathematics, Harvard University, \\
                   One Oxford Street, Cambridge, MA 02318, U.S.A.\\
                   {\it and} \\
                   Jefferson Physical Laboratory, Harvard University,
                   Cambridge, MA 02138, U.S.A.}\\
~\\
\email{sergio.benvenuti@sns.it, sfranco@mit.edu, hanany@mit.edu, 
dario.martelli@cern.ch, sparks@math.harvard.edu}
}

\abstract{We describe an infinite family of quiver gauge theories that are AdS/CFT
dual to a corresponding class of explicit horizon
Sasaki--Einstein manifolds. The quivers may be obtained from a family of
orbifold theories by a simple iterative procedure. 
A key aspect in their construction relies on the global symmetry which
is dual to the isometry of the manifolds. For an arbitrary such
quiver we compute the exact R--charges of the fields in the IR by applying
$a$--maximization. The values we obtain are generically quadratic
irrational numbers and agree perfectly with the central charges and baryon
charges computed from the family of metrics using the AdS/CFT
correspondence. These results open the way for a systematic study of the
quiver gauge theories and their dual geometries.}

\begin{document}


\section{Introduction}

Many lessons have been learned about the dynamics of supersymmetric field theories from their embedding in String 
Theory constructions. Similarly, many properties of the string theory constructions were
revealed by studying the gauge theories embedded in them.

A possible approach to construct interesting gauge theories is by probing singular geometries using D--branes. The geometry 
of the singularity determines the amount of supersymmetry, gauge group, matter content and superpotential interactions on 
the world--volume of the D--branes. Of particular interest are the 4d $\mathcal{N}=1$ gauge theories that arise on a stack 
of D3--branes probing singular Calabi--Yau 3--folds. This reduced amount of supersymmetry allows the possibility of having 
chiral gauge theories. In this paper we will study toric singularities, which represent a relatively simple, yet extremely 
rich, set in the space of possible Calabi--Yau manifolds.
The AdS/CFT correspondence \cite{maldacena} connects the strong coupling regime of gauge theories on D--branes with supergravity 
in a mildly curved geometry. For the case of D3--branes placed at the singularities of metric cones 
over five--dimensional geometries $Y_5$, 
the gravity dual is of the form $AdS_5 \times Y_5$, where $Y_5$ is 
a Sasaki--Einstein manifold \cite{Kehagias,KW,acharya,MP}. 

During the last year, we have witnessed considerable progress in the understanding of these gauge theories. This has 
been due to developments on various different fronts. A key ingredient has been the discovery of the principle of 
$a$--maximization 
\cite{Intriligator:2003jj}, which permits the determination of R--charges of superconformal field theories. This principle is applicable to 
any superconformal field theory, regardless of whether or not 
it is possible to embed the theory in a String Theory construction. 
The $a$--maximization principle has been successively extended in a series of works \cite{Kutasov:2003iy,Intriligator:2003mi,
Kutasov:2003ux,Barnes:2004jj,Kutasov:2004xu}, broadening its range of 
applicability outside of conformal fixed points and bringing us closer to a proof of the supersymmetric $a$--theorem.

Further progress has been made in the study of the non--conformal theories that are produced when, in addition to probe 
D3--branes, fractional D3--branes are included in the system. Fractional D3--branes are D5--branes wrapped on vanishing 
2--cycles of the probed geometry and trigger cascading RG flows, {\it i.e.} flows in which Seiberg duality is used 
every time infinite coupling is reached, generating a sequence of gauge theories, each of them providing a simpler 
description of the theory at every scale. Duality cascades have been studied in detail, and they have been shown to 
exhibit a plethora of interesting phenomena, such as duality walls and chaotic RG flows \cite{Hanany:2003xh,Franco:2003ja,Franco:2003ea,Franco:2004jz}. Recently, supergravity 
duals of cascading RG flows for D3--branes probing complex cones over del Pezzo surfaces have been constructed 
\cite{Franco:2004jz} 
(even without knowledge of the metric of the underlying horizon), validating the applicability of the cascade idea. 
Interesting cascading gauge theories dual to throat geometries with several warp factors (associated to various dynamical 
scales generated by the field theory) can also be studied \cite{Franco:2005fd}. These constructions seem to have potential 
phenomenological applications.

On the geometry side there has also been dramatic progress -- from knowledge of only one non--trivial 
Sasaki--Einstein five--manifold, namely $T^{1,1}$, we now have an infinite family of non--regular
metrics on $S^2 \times S^3$ \cite{paper1,paper2}. These manifolds are called $Y^{p,q}$, where $p$ and $q$ are positive integers with $0 \leq q \leq p$. The associated Type IIB supergravity solutions should be dual to 4d $\mathcal{N}=1$ superconformal field 
theories. 
These theories are superconformal quivers, \emph{i.e.} all the fields trasform in representations of the gauge group with two indices.
From computations using these metrics, it became clear in \cite{paper2} that the dual field theories would exhibit 
very remarkable properties, such as irrational R--charges. The work of \cite{DJ}
has then provided a detailed description of these manifolds and their associated Calabi--Yau singularities
in terms of toric geometry. It turns out that all the cases with $p \leq 2$ are well known and the corresponding superconformal quiver has already been found. $Y^{1,0}$ is the conifold $T^{1,1}$ \cite{KW}. $Y^{2,0}$ is associated to the $\IF_0$ quiver \cite{MP}. The cone over $Y^{1,1}$ is simply $\IC \times \IC^2/\IZ_2$ and the quiver has two gauge groups and $\mathcal{N}=2$ supersymmetry. $Y^{2,1}$, for which the dual quiver gauge theory was computed in \cite{Feng:2000mi} and was also presented in \cite{DJ}, happens to be the first del Pezzo surface (also called $\IF_1$). For this case, the authors of \cite{BBC} have carried out an explicit check of the conformal anomaly coefficient, using $a$--maximisation \cite{Intriligator:2003jj}, finding remarkable agreement\footnote{The main results of \cite{BBC} were computed independently by some of us, unpublished.} with the geometrical prediction of \cite{DJ}. The cone over $Y^{2,2}$ is a $\IZ_4$ orbifold of $\IC^3$, or equivalently a complex cone over the Hirzebruch surface $\IF_2$. In general $Y^{p,p}$ is an orbifold $\IZ_{2p}$ orbifold of $\IC^3$, and the corresponding quiver can be found easily by standard techniques.

The purpose of this paper is to  construct the field theory duals to the entire infinite family of geometries.
Section \ref{The geometries} reviews some properties of the $Y^{p,q}$ geometries. Section $3$ passes to the construction of the associated superconformal quiver gauge theories. The quiver diagrams are constructed and the precise form of the superpotential is found. In general it is a non trivial task to find the exact superpotential. For instance, in the well studied case of del Pezzo singularities the superpotential for del Pezzo $7$ and del Pezzo $8$ is still not known. In the case of the $Y^{p,q}$ manifolds, however, global symmetries and the \emph{quiver toric condition} can be used to fix the complete form of the superpotential. This leads to a successful comparison between global $SU(2) \times U(1)$ flavor symmetries and isometries. Also the $U(1)$ baryonic global symmetry of the theories is shown to follow from the topology of the $Y^{p,q}$ manifolds. From the quiver diagram it is also possible to infer various topological properties of the supersymmetric $3$--cycles of the Sasaki-Einstein manifolds, as we discuss at the end of Section $4$. Here also agreement between gauge theory and geometry is achieved.

Once the quiver diagrams and the exact superpotentials are given, it is a simple exercise to apply the general $a$--maximization procedure of \cite{Intriligator:2003jj}. This leads to a successful comparison between, on the geometry side, volumes of the $5$--manifolds and of some supersymmetric $3$--cycles and, on the gauge theory side, gravitational central charges and R--charges of dibaryon operators.

Having an infinite set of Type IIB solutions, together with their gauge theory duals, represents a substantial 
advancement of our understanding of gauge/gravity duals and opens up the possibility of exciting progress 
in numerous directions. 


\section{The geometries}\label{The geometries}

In this section we give a brief summary of the geometry of the Sasaki--Einstein 
$Y^{p,q}$ 
manifolds, focusing on those aspects which are particularly relevant for 
the construction of, and comparison to, the gauge theory. Further details 
may be found in \cite{paper2, DJ}.

The local form of the $Y^{p,q}$ metrics may be written as
\bea
\label{localmetric}
  \mathrm{d} s^2 &=& \frac{1-y}{6}(\mathrm{d}\theta^2+\sin^2\theta\mathrm{d}\phi^2)+\frac{1}{w(y)q(y)}
      \mathrm{d} y^2+\frac{q(y)}{9}(\mathrm{d}\psi-\cos\theta \mathrm{d}\phi)^2 \nonumber \\ 
      & + &  {w(y)}\left[\mathrm{d}\alpha +f(y) (\mathrm{d}\psi-\cos\theta
      \mathrm{d}\phi)\right]^2\nonumber\\ 
      & \equiv & \mathrm{d} s^2(B)+w(y)[\mathrm{d}\alpha+A]^2
\eea
where
\bea
w(y) & = & \frac{2(b-y^2)}{1-y} \nonumber\\ 
q(y) & = & \frac{b-3y^2+2y^3}{b-y^2} \nonumber\\ 
f(y) & = & \frac{b-2y+y^2}{6(b-y^2)}~.
\eea
Here $b$ is, {\it a priori}, an arbitrary constant\footnote{In \cite{paper2, 
DJ} $b$ was denoted ``$a$''. However, we change notation here to avoid 
any possible confusion with the $a$ central charge of the quivers. 
Both will ultimately have rather similar, but different, expressions in terms of $p$ and $q$.}. These local metrics are Sasaki--Einstein, 
meaning that the metric cone $\mathrm{d} r^2+r^2\mathrm{d} s^2$ is Calabi--Yau. 
For all values of $b$, with $0<b<1$, the base $B$ can be made into a 
smooth manifold of topology $S^2\times S^2$. In particular, the coordinate $y$ 
ranges between the two 
smallest roots $y_1,y_2$ of the cubic $b-3y^2+2y^3$, so $y_1\leq y\leq y_2$.
For completeness we quote the range of the other 4 coordinates: 
$0\leq\theta\leq\pi, 0\leq\phi\leq2\pi, 0\leq\psi\leq2\pi, 0\leq\alpha\leq2\pi\ell$.
Then for a countably infinite number of values of 
$b$ in the interval $(0,1)$ the periods of $\mathrm{d} A$ over the two two--cycles in 
$B$ are rationally related, and hence the metric can be made complete by 
periodically identifying the $\alpha$ coordinate with appropriate 
period. The ratio of the two periods of $\mathrm{d} A$
is then a rational number $p/q$, and by choosing the 
maximal period for $\alpha$ one 
ensures that the Chern numbers $p$ and $q$ for the corresponding $U(1)$ 
principle bundle are coprime. Moreover, 
the bound on $b$ implies that $q<p$.
One now has a 
complete manifold with the topology of a circle fibration over $S^2 \times S^2$. Applying Smale's classification of 5--manifolds, one can deduce 
the topology is always $S^2 \times S^3$. 
For $\mathrm{hcf}(p,q)=h>1$ one 
has a smooth quotient of this by $\IZ_h$. In particular, since 
$H_3(Y^{p,q};\IZ)\cong\IZ\oplus\IZ_h$, the dual 
field theories will possess a baryonic $U(1)_B$ 
flavour symmetry arising from reduction of the Type IIB four--form on the non--trivial non--torsion three--cycle. We denote the complete 
Sasaki--Einstein manifolds obtained in this way by $Y^{p,q}$.

For completeness we give the value of $b$, which crucially determines the cubic 
function appearing in $q(y)$, as well as the two smallest roots $y_1,y_2$ of this 
cubic in terms of $p$ and $q$:
\bea
b & = & \frac{1}{2}-\frac{(p^2-3q^2)}{4p^3}\sqrt{4p^2-3q^2}\nn \\
y_1 & = & \frac{1}{4p}\left(2p-3q-\sqrt{4p^2-3q^2}\right)\nn \\
y_2 & = & \frac{1}{4p}\left(2p+3q-\sqrt{4p^2-3q^2}\right)~.
\eea
The period of $\alpha$ is $2\pi \ell$ where
\beq
\ell = \frac{q}{3q^2-2p^2+p\sqrt{4p^2-3q^2}}\eeq
and the volume is then easily calculated to be
\beq\label{volume}
\mbox{vol}(Y^{p,q})={q^2 [2p+\sqrt{4p^2-3q^2}] \over 3p^2[3q^2-2p^2+p\sqrt{4p^2-3q^2}]} \pi^3~.
\eeq
Notice this is bounded by 
\beq
\mathrm{vol} (T^{1,1}/\IZ_p) > \mathrm{vol} (Y^{p,q}) > \mathrm{vol} (S^5/\IZ_{2}\times \IZ_p) 
\eeq
and is monotonic in $q$. In fact, it will be useful to define 
$Y^{p,0}$ and $Y^{p,p}$ formally as corresponding quotients 
of $T^{1,1}$ and $S^5/\IZ_2$ by $\IZ_p$. These arise naturally as limits
of the toric diagrams for $Y^{p,q}$ \cite{DJ}, although strictly 
speaking the global analysis performed in \cite{paper2} does not 
hold in these limits -- for example, in the case $b=1$ $(p=q)$ the base $B$ 
collapses to a weighted projective space.

It will also be important to recall that these geometries contain two 
supersymmetric submanifolds \cite{DJ}, which are topologically Lens spaces 
$\Sigma_1=S^3/\IZ_{p+q}$ and $\Sigma_2=S^3/\IZ_{p-q}$. 
Here supersymmetric means that the metric cones $C(\Sigma_1)$, $C(\Sigma_2)$ 
are calibrated submanifolds (in fact divisors) in the Calabi--Yau cone. 
These submanifolds 
are located at the two roots $y=y_1$ and $y=y_2$, respectively. 
In fact, the $Y^{p,q}$ manifolds are cohomogeneity one, meaning that the 
isometry group acts with generic orbit of codimension one. The isometry 
group depends 
on $p$ and $q$: for both $p$ and $q$ odd it is $SO(3)\times U(1)\times U(1)$, 
otherwise it is $U(2)\times U(1)$. For a compact cohomogeneity one 
manifold there are then
always precisely two special orbits of higher codimension, and in the 
present case these 
are $\Sigma_1$ and $\Sigma_2$. Note in particular that $SU(2)\sim SO(3)$ 
is contained in the isometry groups.

It is straightforward to 
compute the volumes of $\Sigma_1,\Sigma_2$. However, the following combination
\beq
R[B_i] \equiv \frac{2}{3}\cdot \left(\frac{\pi}{2\mathrm{vol}(Y^{p,q})}\right)\cdot 
\mathrm{vol}(\Sigma_i)\quad\quad i=1,2
\eeq
is more relevant for AdS/CFT purposes, since this formula gives the 
exact R--charges for baryons in the dual gauge theory, arising from 
D3--branes wrapped over the corresponding cycles $\Sigma_i$. These 
are easily calculated \cite{DJ}:
\bea\label{baryons}
R[B_1] & = &
\frac{1}{3q^2}\left[-4p^2+2pq+3q^2+(2p-q)\sqrt{4p^2-3q^2} \right]\nonumber \\
R[B_2] & = & 
\frac{1}{3q^2}\left[-4p^2-2pq+3q^2+(2p+q)\sqrt{4p^2-3q^2} \right]~.
\eea
Note that this formula is homogeneous with respect to re-scaling $p\rightarrow hp, q\rightarrow hq$, implying that manifolds with equal value of the ratio $p/q$ will have the same R--charges.
Let us also note that the R--symmetry in the field theory is dual to the 
canonically defined Killing vector field on the Sasaki--Einstein manifolds
\beq
K  = 3\frac{\partial}{\partial \psi}-\frac{1}{2}
\frac{\partial}{\partial \alpha}~.
\eeq
From the point of view of the Calabi--Yau cone, $K$ arises by contracting 
the Euler vector $r\partial/\partial r$ into the K\"ahler form. 
Note that $K$ has closed orbits precisely when $\ell$ is a rational number, 
since $\psi$ has period $2\pi$ and $\alpha$ has period $2\pi\ell$. 
In this case the Sasaki--Einstein manifold $Y^{p,q}$ 
is said to be quasi--regular, and 
the space of leaves of the foliation defined by $K$ is a K\"ahler--Einstein orbifold. 
This is true if and only if the following quadratic diophantine holds:
\beq
4p^2-3q^2 = n^2~,\eeq
for $n$ an integer number.
If $\ell$ is irrational the generic 
orbits of $K$ do not close, but instead densely fill the
orbits of the torus generated by $[\partial/\partial\psi$,
$\ell\partial/\partial\alpha ]$ and the Sasaki--Einstein manifold $Y^{p,q}$ 
is said to be irregular. 
Note that the orbits close over the 
submanifolds $\Sigma_1$, $\Sigma_2$.

The local form of the metrics is not particularly useful for constructing the 
dual gauge theories. However, one can make contact with the large literature
on gauge theories for D3--branes placed at Calabi--Yau singularities 
by noting  that the group $U(1)^3$ acts as a symmetry of $Y^{p,q}$. 
The Calabi--Yau cone $C(Y^{p,q})$ is thus toric. One can compute the 
toric diagram as follows \cite{DJ}. The K\"ahler form of the Calabi--Yau 
may be regarded as a symplectic form, and one can then introduce a 
moment map for the Hamiltonian torus action by $U(1)^3$, 
which is a map $\mu:C(Y^{p,q})\rightarrow \IR^3$. 
The image is always a polyhedral cone, of a special type, and the 
moment map exhibits the Calabi--Yau as a $U(1)^3$ fibration over this 
polyhedral cone. The latter has four faces, 
where various $U(1)$ subgroups degenerate over the 
faces of the cone in a way determined by the normal vectors to the faces. 
One can now apply a Delzant theorem to this cone to 
obtain a gauged linear sigma model for $C(Y^{p,q})$. This is a simple
algorithm that takes the combinatorial data that defines the 
polyhedral cone and produces the charges of the gauged linear 
sigma model. The result  \cite{DJ} is a $U(1)$ theory with 4 chiral superfields 
with charges $(p,p,-p+q,-p-q)$. Equivalently, because the 
space we start with is Calabi--Yau, the normal vectors to the four faces 
of the polyhedral cone lie in a plane. Projecting the four 
vectors onto this plane gives the vertices of the toric diagram.

\section{The quiver theories}\label{sec:quivers}

In this section we will present the quiver theories for the infinite class of manifolds which were presented 
in the previous section. We will recall the toric diagrams of each manifold, draw its corresponding (p,q)--web, 
extract simple information like the number of nodes and the number of fields in a given quiver theory from its 
corresponding toric diagram, and then present the quiver itself. Finally we write down the superpotential for 
the quiver theory.

\begin{figure}[ht]
  \epsfxsize = 4.5cm
  \centerline{\epsfbox{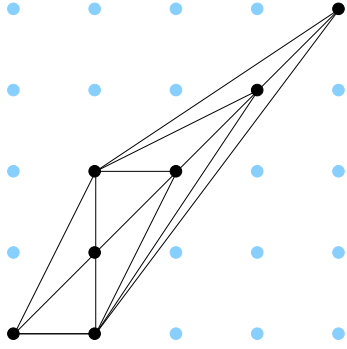}}
  \caption{Triangulation of the toric diagram for $Y^{4,2}$. The number of gauge groups in the associated
        quiver theory is given by the number of triangles, which in this case is equal to eight.}
  \label{triangulation_toric}
\end{figure}

\begin{figure}[ht]
  \epsfxsize = 13cm
  \centerline{\epsfbox{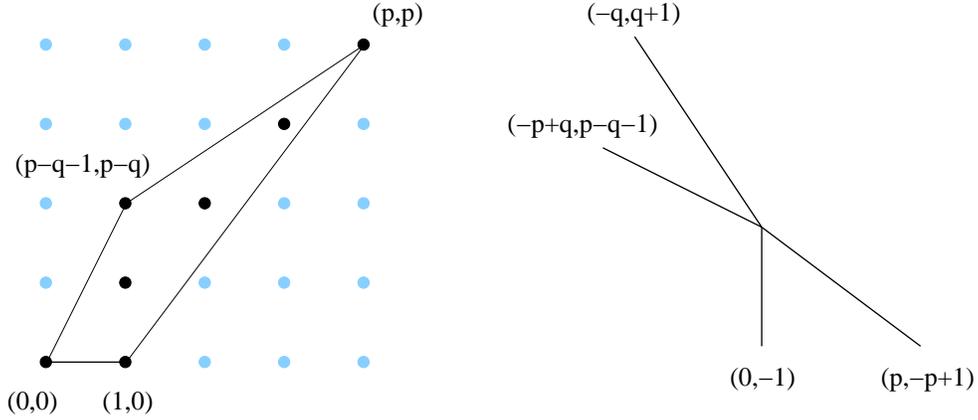}}
  \caption{Toric diagram and external legs of the corresponding (p,q)-web for $Y^{4,2}$.}
  \label{web_general}
\end{figure}

The toric diagrams for $Y^{p,q}$ were found in \cite{DJ} and are defined by the convex polygon over a $\IZ^2$ lattice
defined by four vertices located at
\beq
[0,0],\qquad[1,0],\qquad[p,p],\qquad[p-q-1,p-q]~.
\label{vertices_toric}
\eeq
See \fref{triangulation_toric} for the toric diagram of $Y^{4,2}$ and \fref{web_general} for a schematic description of 
the general case. Given the toric diagram, it is in principle possible to determine the gauge theory for any $Y^{p,q}$, by 
the process of partial resolution \cite{Feng:2000mi}. The starting point can be for example an abelian orbifold of the
form $\IC^3 / \IZ_m \times \IZ_n$, with $m$ and $n$ sufficiently large. In particular, the toric 
diagrams of $Y^{p,q}$ could be obtained by partial resolutions of the 
$\IC^3 / \IZ_{p+1}\times \IZ_{p+1}$ orbifold, for any $q$ at fixed $p$. 
Partial resolution corresponds to turning on non--generic
Fayet--Illiopoulos parameters that reduce the toric diagram to the desired one. This method becomes computationally 
intractable even for modest values of $p$ and $q$, and thus different approaches have to be developed.

We would like to get as much information as possible about the gauge theory from this toric description. Given a toric diagram, there are three steps in determining a supersymmetric quiver gauge theory with 4 supercharges that is associated to it.
First we would like to get the number of gauge groups. Second we look for the number of bifundamental fields and the gauge 
quantum numbers. Finally we find which of the allowed gauge invariant terms appear in the superpotential. We will see now
that, using very simple geometric ideas, it is possible to go far in answering the first two questions.

For a given toric diagram the number of gauge groups is a constant associated to the geometry.
It is independent of any action of dualities that the gauge theory is undergoing.
One way to look at the different gauge groups of the quiver is as living on the world volume of fractional branes which are given by bound states of D--branes which wrap even dimensional cycles in the geometry.
These are the number of possible ways in which D--branes (3, 5, and 7--branes) 
can be wrapped on 0, 2 and 4--cycles, respectively. For the manifolds under study this number turns out to be particularly simple and is just the Euler characteristic of the 4d--base. In the toric diagram this number is given by the number
of triangles in any possible triangulation of the corresponding diagram. Equivalently the number of triangles is given by the area of the toric diagram in units in which a single triangle has area 1. Different triangulations
are related by flops, which correspond in the gauge theory to Seiberg duality transformations. 
Let us first notice that the vertex $(p-q-1,p-q)=(p-q,p-q)+(-1,0)$ sits always on a line parallel to the one
joining the $(0,0)$ and $(p,p)$ points, located one lattice spacing to the left of it. 
In order to count the number of triangles, we can use a uniform triangulation for every $Y^{p,q}$, given by
the line that joins the points $(0,0)$ and $(p,p)$, and the segments that connect
$(1,0)$ and $(p-q-1,p-q)$ with the $(i,i)$ points for $0 \leq i \leq p$. It is clear from this construction 
that the quiver associated to $Y^{p,q}$ has $2p$ gauge groups, namely $2p$ nodes. We illustrate this triangulation
in \fref{triangulation_toric} for the example of $Y^{4,2}$.

Every toric diagram has an associated (p,q)--web, corresponding to the reciprocal diagram in which lines are replaced by orthogonal lines and nodes are exchanged with faces. The boundary 
of the toric diagram determines the charges of the external legs of the web. \fref{web_general} shows this construction for the case of $Y^{4,2}$ 
\footnote{The cones over $Y^{p,q}$ are generically examples of geometries with more than one collapsing 4--cycle. The
study of the gauge theories using (p,q)--webs was initiated in \cite{Franco:2004wp}, in the context of quivers obtained
by general Picard--Lefschetz monodromies.}. 

Furthermore, external legs determine the total number of bifundamental fields using the formula

\beq
n_{fields}={1\over 2}\sum_{i,j\in \mbox{legs}}^4 \left| \det \left( \begin{array}{cc} p_i & q_i \\
p_j & q_j \end{array}\right) \right|~.
\label{nfields}
\eeq

This comes from the mapping of 0, 2 and 4--cycles to 3--cycles in the mirror manifold and computing their intersection,
as described in \cite{Hanany:2001py}.

For $Y^{p,q}$, the charges of the external legs of the web diagram can be computed from the toric diagram given
by \eref{vertices_toric}, and are

\beq
\begin{array}{l}
(p_1,q_1)=(-p+q,p-q-1) \\
(p_2,q_2)=(-q,q+1) \\
(p_3,q_3)=(p,-p+1) \\
(p_4,q_4)=(0,-1)
\end{array}
\label{pqcharges}
\eeq
from which, using \eref{nfields}, we can compute $n_{fields}=4p+2q$.

The determination of the superpotential typically is the most difficult task in completing the quiver theory and at the 
moment we do not have a general method of computing it for an arbitrary toric diagram. However, an important restriction 
for any quiver theory corresponding to an affine toric variety is that each of the fields in the quiver appears in the 
superpotential precisely twice ({\it i.e.} the  F--term equations are of the form monomial equals monomial). As a result when 
counting the total number of fields appearing in each of the polygons contributing to the superpotential we 
should find $8p+4q$ such fields. 

In addition, geometric blow--downs correspond to Higgsings in the gauge theory.
In such cases, the non--zero expectation value of a bifundamental field introduces a scale. When running the RG flow
to scales much smaller than this vev, one encounters the gauge theory for the blown--down geometry. This approach
has been implemented in \cite{Feng:2002fv} to derive the superpotentials of several gauge theories. Furthermore, the (p,q)--web representation of the toric singularities enables a simple identification of the bifundamental field acquiring
a vev \cite{Franco:2002ae}. It turns out that the $Y^{p,q}$ geometries 
can be  blow--down to the 
$\IC^3 / \IZ_{p+q}$ orbifold, for which the quiver and the superpotential
are known by standard methods. It is then possible to perform a further check 
of our construction 
by verifying that the proposed superpotential 
produces the final gauge theory after Higgsing.


In the case at hand, as explained in the previous section, the superconformal field theories we are looking for possess a $SU(2)$ global symmetry. This considerably restricts the possible choices of superpotential and, combining this requirement with the toric restrictions (each field has to appear exactly twice in the superpotential, one time with sign \emph{plus} and one time with sign \emph{minus}), it will turn out that in all of the cases there is precisely one superpotential satisfying all the properties, modulo an overall rescaling.


\subsection{An iterative procedure starting with the $Y^{p,p}$ quiver}

We now move on and construct the quiver for $Y^{p,q}$. That is, we will now determine how the $4p+2q$ bifundamental 
fields of $Y^{p,q}$ are charged under its $2p$ gauge groups.

A convenient way to construct the quiver theories for the $Y^{p,q}$ manifolds for a fixed $p$ is to start with the case
 $q=p$ and 
work our way down. For the case $q=p$, $Y^{p,p}$ is the base of the orbifold $\IC^3/\IZ_{2p}$. This orbifold group has 
an action on the three coordinates of $\IC^3$, $z_i, i=1,2,3$ by $z_i\rightarrow\omega^{a_i}z_i$ with $\omega$ a $2p$--th 
root of unity, $\omega^{2p}=1$, and $(a_1,a_2,a_3)=(1,1,-2).$ Since $2p$ is always even the group $\IZ_{2p}$ with this 
action is actually reducible and one can write this group action as $\IZ_2\times\IZ_p$. For special cases of $p$ the 
group $\IZ_p$ can be further reducible with the induced action but to keep the discussion general we will just refer to 
this group as $\IZ_p$, without looking at the detailed structure, and bearing in mind that this group can be reducible.

\begin{figure}[ht]
  \epsfxsize = 6cm
  \centerline{\epsfbox{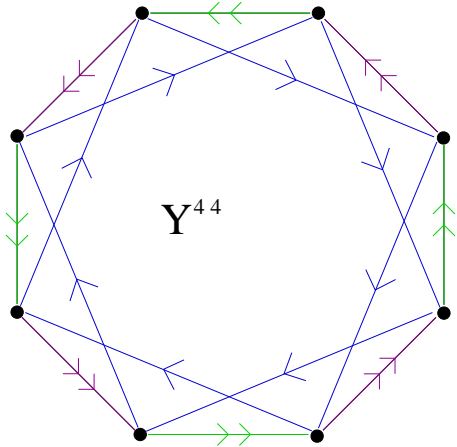}}
  \caption{Quiver diagram for the $\IC^3/\IZ_8$ orbifold. We have color--coded bifundamental fields in accordance to
  the forthcoming discussion. Superpotential terms appear in the quiver diagram as triangles combining a green, a blue 
  and a cyan arrow.}
  \label{quiver_Z8}
\end{figure}

The quiver theory for an orbifold is particularly simple and can be computed along the lines given in 
\cite{Hanany:1998sd}. As stated above it has $2p$ nodes and, using the formula below equation (\ref{pqcharges}) for 
the number of fields, we find that there are $6p$ bifundamental fields. \fref{quiver_Z8} shows the quiver theory
for the $\IC^3/\IZ_{8}$ orbifold.

Since $a_1=a_2$ there is a natural $SU(2)\times U(1)$ isometry of this space. The $SU(2)$ part acts on the 
coordinates $z_1$ and $z_2$ which transform as a doublet and the $U(1)$ part acts as the subgroup of $SU(3)$ which 
commutes with this $SU(2)$. This isometry becomes a global symmetry of the quiver gauge theory. All fields and their 
combinations transform in an irreducible representation of this group. As a result we can divide the $6p$ fields 
into $2p$ doublets that go along the edges of a polygon of $2p$ nodes. There are additional $2p$ singlet fields, which
form triangles with each of two adjacent edges. There are $2p$ such triangles, all of which 
contribute to the superpotential. We require invariance of the theory under the global symmetry and therefore each 
time we have two doublets in the superpotential it should be understood that they are contracted by an epsilon symbol 
and therefore there will be two terms for each such polygon. We can now count the number of fields in the superpotential 
to be $2p\cdot3\cdot2=12p$, as expected from the fact that this quiver corresponds to an affine toric variety. Specifically 
we denote the doublet fields as $X_{i}^\alpha, i=1,\dots,2p, \alpha=1,2$, with $i$ labeling the $\IZ_{2p}$ index which takes values mod $2p$, 
while $\alpha$ labels the $SU(2)$ global symmetry index. Furthermore, we denote the singlets as $Y_i, i=1,\ldots,2p$. 
We use the convention that an arrow is labeled by the node number at which it starts. The superpotential then takes the 
simple form
\beq
W=\sum_{i=1}^{2p}\epsilon_{\alpha\beta}X_i^\alpha X_{i+1}^\beta Y_{i+2}.
\eeq
It is understood in this notation that the gauge quantum numbers are summed over in cyclic order and are therefore 
suppressed. In what follows, and due to the fact the the orbifold group $\IZ_{2p}$ is reducible to at least $\IZ_2\times\IZ_p$, it will be convenient to rename the $X$ fields as follows: $U_i=X_{2i}, V_i=X_{2i+1}$. Note that the fields $U$ are even under $\IZ_2$ while the fields $V$ are odd under $\IZ_2$. From now
on we will adhere to the convention, already used in \fref{quiver_Z8}, of indicating $V_i$ fields in green and $U_i$ 
fields in cyan. In terms of these fields the superpotential takes the form
\beq
W=\sum_{i=1}^{p}\epsilon_{\alpha\beta}(U_i^\alpha V_{i}^\beta Y_{2i+2}+V_i^\alpha U_{i+1}^\beta Y_{2i+3}).
\eeq

The gauge theory for $Y^{p,p-1}$ results from the following set of operations, which remove
three fields and add one:

\begin{itemize}
\item Pick an edge of the polygon, say the one which has an arrow $V_i$ starting at node $2i+1$, and remove one arrow from the corresponding doublet to make it a singlet. Call this type of singlet $Z_i$.

\item Remove the two diagonal singlets, $Y$ that are connected to the two ends of this singlet $Z$. Since we chose the $V_i$ arrow which is removed to start at node $2i+1$ the $Y$ fields which are removed are $Y_{2i+2}$ and $Y_{2i+3}$.
This action removes from the superpotential the corresponding two cubic terms that involve these $Y$ fields.

\item Add a new singlet $Y_{2i+3}$ in such a way that, together with
the two doublets at both sides of the singlet $Z_i$, they form a rectangle. Specifically this arrow starts at node $2i+3$ and ends at node $2i$. The new rectangle thus formed contains two doublets which as before should be contracted to an $SU(2)$ singlet. This term is added to the 
superpotential.

\end{itemize}

\begin{figure}[ht]
  \epsfxsize = 6cm
  \centerline{\epsfbox{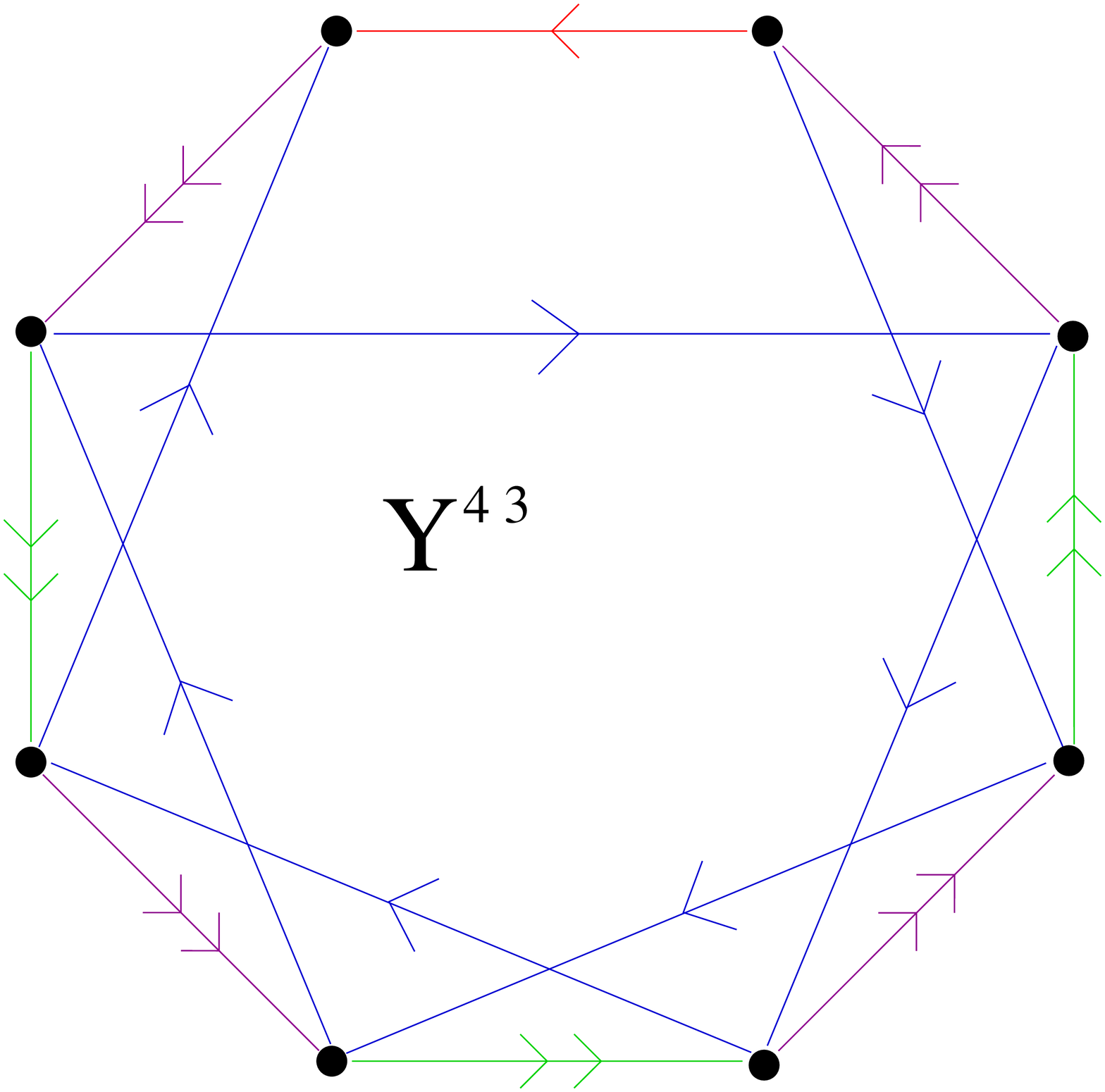}}
  \caption{Quiver diagram for $Y^{4,3}$, obtained from $Y^{4,4}=\IC^3/\IZ_8$.}
  \label{quiver_43}
\end{figure}

By the end of this process, we get $6p-2$ fields. There are $p$ doublet fields $U_i$, $p-1$ doublet fields 
$V_j, j\not=i$, one field of type $Z_i$ and $2p-1$ diagonal singlets of type $Y_j, j\not=2i+2$. We present in
\fref{quiver_43} the $Y^{4,3}$ example, obtained from $Y^{4,4}=\IC^3/\IZ_8$ by the series of steps outlined
above. We indicate the new $Z$ singlet in red.

The new superpotential has $(2p-2)$ triangles and $1$ rectangle (recall that when we refer to one
triangle or one rectangle, we are actually indicating the $SU(2)$ invariant combination given by two of them). 
The resulting superpotential is
\beq
W=\sum_{i\not=j=1}^{p}\epsilon_{\alpha\beta}(U_j^\alpha V_{j}^\beta Y_{2j+2}+V_j^\alpha U_{j+1}^\beta Y_{2j+3})+\epsilon_{\alpha\beta}Z_iU_{i+1}^\alpha Y_{2i+3}U_i^\beta.
\eeq
As a check we can verify that the model still satisfies the toric condition regarding the number of fields in the 
superpotential. There are $(p-1)\cdot3\cdot2\cdot2+4\cdot2=12p-4$, in agreement with our expectation.

\begin{figure}[ht]
  \epsfxsize = 6cm
  \centerline{\epsfbox{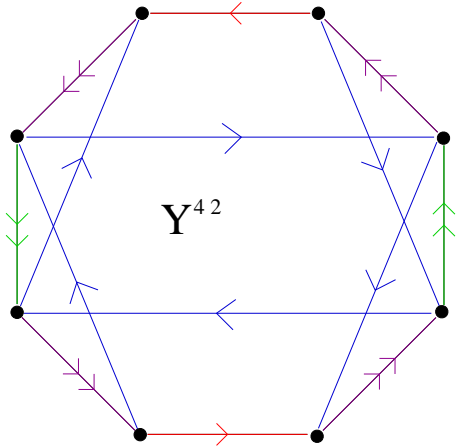}}
  \caption{Quiver diagram for $Y^{4,2}$, obtained from $Y^{4,4}=\IC^3/\IZ_8$ by applying the three step sequence twice.}
  \label{quiver_42}
\end{figure}

We now continue to construct the $Y^{p,p-2}$ model. This is an easy task and is just a repetition of the 3--step 
process described above. We pick an index $j\not=i$ and turn a $V_j$ doublet into a $Z_j$ singlet by
repeating the sequence of steps previously explained.
The result is a theory with $6p-4$ fields forming $p$
$U$ doublets, $p-2$ $V$ doublets, $2$ $Z$ singlets and $2p-2$ $Y$ singlets.
We present the $Y^{4,2}$ example in \fref{quiver_42}.

\begin{figure}[ht]
  \epsfxsize = 6cm
  \centerline{\epsfbox{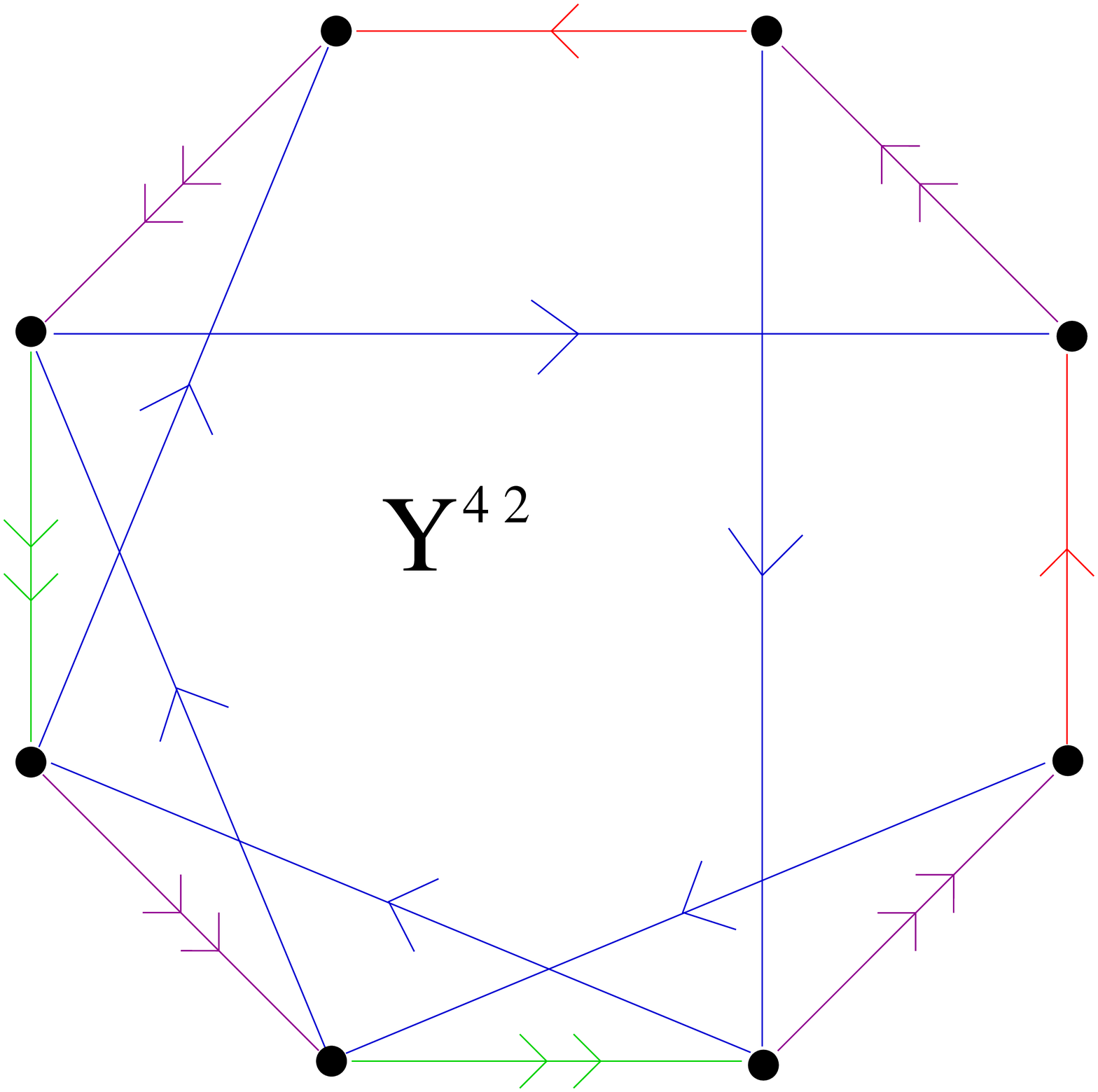}}
  \caption{A different quiver diagram for $Y^{4,2}$, corresponding to a different toric phase.}
  \label{quiver_42b}
\end{figure}

When one applies the procedure the second time, there is the possibility of choosing the double leg to ``open up''. 
For instance in the case of $Y^{4,2}$ there are two different choices that can be made. One is \fref{quiver_42}, 
the other is \fref{quiver_42b}.

These two quivers are different, but are actually related by Seiberg duality. They correspond to two different 
``toric phases'' of the same Duality Tree \cite{Feng:2002zw,Franco:2003ja,Benv-Hanany}.

The superpotential now has $2p-4$ triangles and $2$ rectangles and is given by
\beq
W=\sum_{i\not=j\not=k=1}^{p}\epsilon_{\alpha\beta}(U_k^\alpha V_{k}^\beta Y_{2k+2}+V_k^\alpha U_{k+1}^\beta Y_{2k+3})+\epsilon_{\alpha\beta}\sum_{k=i,j}Z_kU_{k+1}^\alpha Y_{2k+3}U_k^\beta
\eeq
and has $(p-2)\cdot3\cdot2\cdot2+2\cdot4\cdot2=12p-8$ fields, which is consistent with the fact that the probed geometry
is toric.

We can keep going down in $q$ by iterating the procedure above.
Thus, for $Y^{p,q}$ there are $4p+2q$ fields forming $p$ $U$ doublets, $q$ $V$ doublets,
$(p-q)$ $Z$ singlets and $(p+q)$ diagonal singlets $Y$. The superpotential has $2q$
triangles and $(p-q)$ rectangles. The general superpotential is
\beq
W=\sum_{k}\epsilon_{\alpha\beta}(U_k^\alpha V_{k}^\beta Y_{2k+2}+V_k^\alpha U_{k+1}^\beta Y_{2k+3})+\epsilon_{\alpha\beta}\sum_{k}Z_kU_{k+1}^\alpha Y_{2k+3}U_k^\beta.
\eeq
The sum $k$ for the cubic terms is in indices in which $V$ exists and the sum $k$ for the quartic terms is 
in indices in which $Z$ exists. Note that any of the indices 1 to $p$ appears precisely once either in the cubic or the quartic sum. The number of fields in the superpotential is $q\cdot3\cdot2\cdot2+(p-q)\cdot4\cdot2=8p+4q$, verifying the quiver is toric.

For completeness we also give the quivers for $Y^{4,1}$ and $Y^{4,0}$.
\begin{figure}[ht]
  \epsfxsize = 6cm
  \centerline{\epsfbox{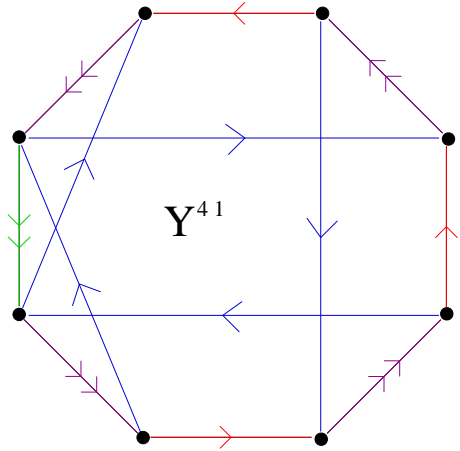}}
  \caption{Quiver diagram for $Y^{4,1}$. In this case we just see one toric phase.}
  \label{quiver_41}
\end{figure}

\begin{figure}[ht]
  \epsfxsize = 6cm
  \centerline{\epsfbox{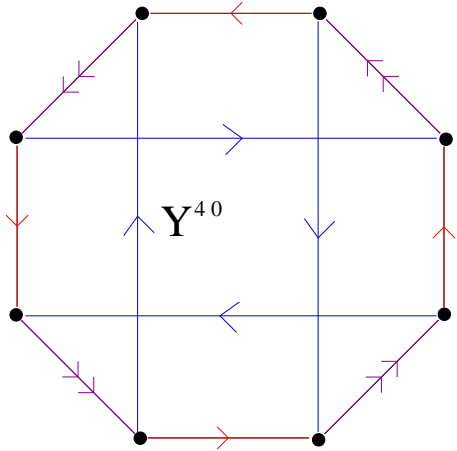}}
  \caption{Quiver diagram for $Y^{4,0}$. Note that the superpotential terms are only quartic. Correspondingly, the nodes have precisely $2$ incoming and $2$ outgoing arrows. This quiver diagram is indeed a $\IZ_4$ orbifold of the conifold.}
  \label{quiver_40}
\end{figure}

All the different quivers constructed by our iterative procedure satisfy the following property. In the $Y^{p,p}$ models every node has precisely $3$ incoming and $3$ outgoing arrows. Each time the procedure is applied, for precisely two nodes of the quiver the number of incoming and outgoing arrows becomes $2$. At the end of the process we are left with a quiver where all of the nodes have precisely $2$ incoming and $2$ outgoing arrows. A way to rephrase this fact is by saying that the ``relative number of flavors'' for each gauge group passes from $3$ to $2$. This ``relative number of flavors'' is discussed in detail in \cite{Benv-Hanany} and is useful in order to understand the structure of Seiberg dualities for any superconformal quiver. The whole set of Seiberg dual phases of the same theory can be organized in a Duality Tree \cite{Franco:2003ja, Franco:2003ea}. Also for the models we are discussing it is generically possible to construct an infinite tower of superconformal quivers related to the ones constructed here by applying Seiberg dualities. The fact that the relative number of flavors is always greater than (or equal to) $2$ implies that we are just seeing the ``minimal models'' of the Duality Tree \cite{Benv-Hanany}. We notice that a generic quiver in the Duality Tree will not satisfy the quiver toric condition, {\it i.e.} the equality for all the ranks of the gauge groups. In many cases it is known that the different models in the Duality Tree are classified by solutions of Diophantine equations; it would be nice to understand if this is true also here.

Closed loops in the Duality Tree can be used to engineer Duality Cascades of  Klebanov--Strassler type. In the conifold case however the Duality Tree is trivial, meaning that it is composed of just one theory and there is just one closed loop of length one. We thus expect interesting generalisations of the Duality Cascade to be found here, such as ``Duality Walls'' \cite{Hanany:2003xh,Franco:2003ja, Franco:2003ea}.

\subsection{Higgsing the $Y^{p,q}$ quivers}
In some special cases, the $Y^{p,q}$'s correspond to geometries whose associated
gauge theories are well understood. We have already seen that $Y^{p,p}$ corresponds to $\IC^3/\IZ_{2p}$. In addition,
$Y^{p,0}$ has no triangles at all and the R--charges for bifundamental
fields are $1/2$. This agrees with the fact that, as is clear from the corresponding toric diagram, 
$Y^{p,0}$ corresponds to the $\IZ_p$ orbifold of the conifold.

Another appealing observation that follows from our construction of the general quiver is that the quiver for $Y^{p,q}$ can be Higgsed to the one for the orbifold 
$\IC^3 / \IZ_{p + q}$ by turning on non--zero vevs for all the $(p - q)$ $Z$ fields.
One can see this geometrically as follows. We begin with
$S^5$, viewed as the unit sphere in $\C^3$ with complex
coordinates $(z_1,z_2,z_3)$. Consider the $U(1)$ action with weights
$(1,1,-2)$, so that $(z_1,z_2,z_3)\mapsto (\lambda z_1, \lambda z_2,
\lambda^{-2}z_3)$ with $\lambda \in U(1)$. The quotient by this 
action is a form of weighted
projective space, which we denote $W\C P^2_{[1,1,-2]}$. However, before we
quotient out, we may first factor through by $\IZ_{p+q}\subset U(1)$ to 
give $S^5/\IZ_{p+q}$. This is precisely the orbifold that is dual 
to the Higgsing described above.

A useful description of $W\C P^2_{[1,1,-2]}$ is as follows.
One takes $T^*S^2$, which has boundary $\IR P^3 = S^3/\IZ_2$, 
and glues onto this the $A_1$ singularity
$\IR^4/\IZ_2$. In this realisation the two--sphere
$(z_1,z_2,0)$ corresponds to the zero section of $T^*S^2$ whereas the
$A_1$ singularity is located at the point $(0,0,z_3)$. The idea now 
is to blow up the $A_1$ singularity in the base in
the usual way, replacing it with another copy of $T^*S^2$.
The resulting space is an $S^2$ bundle over
$S^2$ in which the gluing function across the equator
corresponds to $2\in \IZ\cong \pi_1(U(1))$, where
$U(1)\subset SO(3)$ acts on the fibre two--sphere. This bundle can also be
made by gluing $T^*S^2$ to minus itself along the common boundary
$\IR P^3$. Notice that this is precisely the topological construction 
of the base $B$ of the Sasaki--Einstein manifolds $Y^{p,q}$ in \cite{paper2}.

Having resolved the base, we must now consider the fibre $S^1$. 
Notice that over $W\C P^2_{[1,1,-2]}$ minus its
singular point, which gives topologically $T^*S^2$, the
original $U(1)$ bundle has winding number $p+q$ over $S^2$. 
However, note that
$H^2(\IR P^3;\IZ)\cong \IZ_2$. One easily sees
that the map from $\IZ$,
which determines the topology of the $U(1)$ bundle over $T^*S^2$, 
to $\IZ_2$, which determines the topology on the
boundary $\IR P^3$, is just reduction modulo 2. To extend the 
$U(1)$ bundle over the blown--up copy of  $S^2$, 
topologically we must specify an integer $l\in \IZ$
which gives the winding number over the blown--up cycle. However,
in order for this to glue onto the existing $U(1)$ bundle
described above, it is clear that we must have $l\cong {p+q}$ mod 2 in
order that the boundaries match. The resulting space is a $U(1)$
bundle over $B$ with winding numbers $p+q$ and
$l$ over two $S^2$ zero sections. Note that these were called $S_1,S_2$ in 
\cite{paper2} and \cite{DJ}. Moreover, without loss of generality we 
may set $l=p-q$, since $l\cong p+q$ mod 2.

Notice that the final space has precisely the topology of $Y^{p,q}$. 
Moreover, we also have the following relation between 
volumes:
\be \vol(S^5/\IZ_{2p})<\vol(Y^{p,q})<\vol(S^5/\IZ_{p+q})\quad \quad q<p~.\ee
This process we have described is therefore 
consistent with an $a$--theorem for Higgsing.

As an example consider the model $Y^{4,3}$; giving a vev to the bifundamental field $Z$ and flowing to the infra--red
there is a Higgsing mechanism. The gauge group passes from $U(N)^8$ to $U(N)^7$ and the quartic term in the superpotential disappears. In summary, the low energy theory becomes the known orbifold $\IC^3 / \IZ_{7}$.

The same procedure can be applied to one of the phases of $Y^{4,2}$. Here new features arise. Giving a vev to both of the 
two $Z$ fields one ends up with the orbifold $\IC^3 / \IZ_{6}$, which is actually the model $Y^{3,3}$. This fact also 
relates some observations made in Section 2 about monotonic behaviour of the volumes of $Y^{p,q}$ with the supersymmetric 
$a$--theorem.

Giving instead a vev for just one of the $Z$ fields one finds a new model, which is the orbifold $\IC^3 / \IZ_{7}$ where 
the above described three--step operation has been applied. These types of models are not part of the $Y^{p,q}$ series, 
as can be seen from the fact that all the $Y^{p,q}$ models have precisely one baryonic $U(1)$ symmetry. The quivers 
$\IC^3 / \IZ_{odd}$ instead cannot have precisely one $U(1)$ baryonic symmetry, since the number of baryonic symmetries 
is given by the number of nodes of the quiver minus the rank of the (antisymmetric part of) the quiver intersection matrix 
minus one. Since an antisymmetric matrix always has even rank, in a quiver with an odd number of nodes the number of 
baryonic $U(1)$'s is always even.

\section{R--charges and horizon volumes}

In this section we compute the exact R--charges as well as the $a$ central charge
of the $Y^{p,q}$ quivers using $a$--maximization, and compare with  the geometrical predictions
of \cite{paper2} and \cite{DJ}.  The agreement found is perfect.

Let us first recall the logic of $a$--maximization. As explained in \cite{Intriligator:2003jj},
in this procedure one assigns some trial R--charges to the different fields and the
\emph{exact} R--charges are then determined by those values that (locally) maximize
the combination of 't Hooft anomalies found in \cite{Anselmi:1998am,Anselmi:1998ys}:
\begin{equation}
a(R)= \frac{3}{32}(3\mathrm{tr} R^3 -  \mathrm{tr} R)~.
\end{equation}
The maximal value of this function is then precisely the exact $a$ central charge
of the theory at the IR fixed point. As proposed in \cite{Intriligator:2003jj} the trial R--charges can be chosen
by assigning a fiducial R--charge $R_0$, provided the latter satisfies the constraints imposed by
anomaly cancellation. The fiducial R--charge is allowed to mix with the abelian global symmetries, which
for all the $Y^{p,q}$ quivers is $U(1)_F \times U(1)_B$.  

We find it more convenient to implement this procedure in the following equivalent 
fashion \cite{Franco:2003ja,Herzog:2003dj}. Recall that for a supersymmetric gauge theory with gauge group
$G$, the beta function for the gauge coupling $\alpha =g^2/4\pi$ is
\begin{equation}
\beta(\alpha) =
-\frac{\alpha^2}{2\pi}\frac{3T(G)-\sum_iT(r_i)(1-\gamma_i(\alpha))}{1-\frac{\alpha}{2\pi}T(G)}
\end{equation}
where $\gamma_i$ is the anomalous dimension of a chiral superfield in the representation $r_i$,
and for $G=SU(N)$ the Casimirs take the values $T(\mathrm{fund})=1/2$ and $T(G)=T(\mathrm{adj})=N$.

At the IR fixed point all the numerators of the beta functions corresponding to each gauge group factor (node of the quiver) must vanish, thus imposing the relations
\begin{equation}
N -\frac{1}{2}N\sum_i (1-R_i)=0\label{betavanish}
\end{equation}
where we used the fact that at the fixed
point the anomalous dimension is related to the R--charge as $\gamma_i=3R_i-2$. We have also used the fact that our quivers are always ``toric'', in the sense that all the ranks of the gauge groups are equal (to $N$ in this case).
We can then consider a set of arbitrary R--charges $R_i$ which satisfy equation
(\ref{betavanish}) at each node, as well as the additional requirement that each monomial in the superpotential
has R--charge precisely 2.

Let us illustrate this procedure with an example, and
then move on to the general $Y^{p,q}$ quivers.

\subsection{Gauge theory analysis for $Y^{3,2}$}

At the bottom of the infinite family we
have $Y^{2,1}$, which is a metric on the horizon of the complex cone over 
$dP_1$ \cite{DJ}. 
The gauge theory for this geometry was computed in \cite{Feng:2000mi} and 
is also presented in \cite{Feng:2002zw}. It has recently been discussed 
in reference \cite{BBC}, resolving an apparent mismatch between 
gauge theory results in the literature and the geometric analysis of \cite{DJ}.
The next case is $Y^{3,2}$ and the corresponding 
quiver is presented in Figure \ref{quiver32fig} below.
\begin{figure}[!th]
\vspace{0mm}
\begin{center}
\epsfig{file=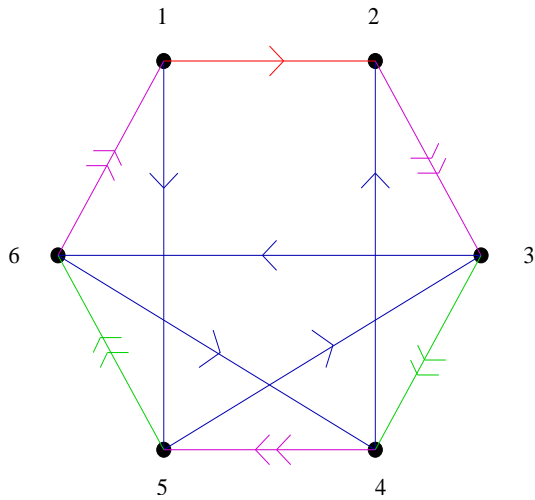,width=7cm,height=6.5cm}\\
\end{center}
\caption{Quiver diagram for $Y^{3,2}$.} 
\label{quiver32fig} \vspace{0mm}
\end{figure}
%
The gauge group for the theory is $U(N)^6$. We may now determine the exact central charge of this theory
 using $a$--maximization.
For the $Y^{3,2}$ quiver we have $1+5+5=11$ {\it a priori} different R--charges, subject to
$6$ linear relations coming from (\ref{betavanish})
\begin{small}
\bea
\frac{\beta_1}{N} & = & 1 +\frac{1}{2} (R_{12}-1)+\frac{1}{2}(R_{15}-1)+\frac{1}{2}2(R_{61}-1) = 0 \nn\\
\frac{\beta_2}{N} & = & 1 +\frac{1}{2} (R_{12}-1)+\frac{1}{2}(R_{42}-1)+\frac{1}{2}2(R_{23}-1) = 0 \nn\\  
\frac{\beta_3}{N} & = & 1 +\frac{1}{2} (R_{53}-1)+\frac{1}{2}(R_{36}-1)+\frac{1}{2}2(R_{23}-1)
+\frac{1}{2}2(R_{34}-1)= 0 \nn\\
\frac{\beta_4}{N} & = & 1 +\frac{1}{2} (R_{64}-1)+\frac{1}{2}(R_{42}-1)+\frac{1}{2}2(R_{34}-1)
+\frac{1}{2}2(R_{45}-1)= 0 \nn\\
\frac{\beta_5}{N} & = & 1 +\frac{1}{2} (R_{15}-1)+\frac{1}{2}(R_{53}-1)+\frac{1}{2}2(R_{45}-1)
+\frac{1}{2}2(R_{56}-1)= 0 \nn\\
\frac{\beta_6}{N} & = & 1 +\frac{1}{2} (R_{36}-1)+\frac{1}{2}(R_{64}-1)+\frac{1}{2}2(R_{56}-1)
+\frac{1}{2}2(R_{61}-1)= 0 
\eea
\end{small}
and $5$ conditions from the superpotential 
\begin{small}
\bea
R_{56}+R_{61}+R_{15} & = & 2\nn\\
R_{45}+R_{56}+R_{64} & = & 2\nn\\
R_{34}+R_{45}+R_{53} & = & 2\nn\\
R_{23}+R_{34}+R_{42} & = & 2\nn\\
R_{12}+R_{23}+R_{36}+ R_{61} & = & 2~.
\eea
\end{small}
However, one can check that two charges remain undetermined -- this is a general feature, valid for all the $Y^{p,q}$ models, and is related to the fact that the global symmetry is always $U(1) \times U(1)$. The 
maximization is then always performed over a two dimensional space of trial R--charges.

We can parameterize the two trial R--charges as follows:
\bea
R_{12}=x\qquad R_{36}=R_{15}=R_{64}=R_{53}=R_{42}=y\nn\\
R_{34}=R_{56}=1+\frac{1}{2}(x-y)\qquad R_{61}=R_{45}=R_{23}=1-\frac{1}{2}(x+y)~.
\eea
Recall the definition of $a$ in terms of these R--charges:
\be
a  = \frac{3}{32} \big(2|G|+\sum_i3(R_i-1)^3-(R_i-1)\big)
\ee
where $|G|$ is the number of vector multiplets.
Here it is straightforward to check that tr$ R=0$ as shown on general grounds in \cite{Benv-Hanany}. 
One  can now compute $a(x,y)= 9/32 \mathrm{tr} R^3$ which reads\footnote{Here, and henceforth, we suppress
factors of $N$.}
\bea
\frac{32}{9}a(x,y)& = & 6 +(x-1)^3+ 5 (y-1)^3 +\frac{1}{2} (x-y)^3-\frac{3}{4}(x+y)^3 ~.
\eea
The local maximimum is found at
\bea
x_{max}  =  \frac{1}{3} (-9+4\sqrt{6}) \qquad y_{max}  =  -1 + 2\sqrt{\frac{2}{3}}
\eea
for which  we find $a_{max} = \frac{27}{16} (-9+4 \sqrt{6})$, which indeed
agrees with $\pi^3/(4\cdot \mathrm{vol}(Y^{3,2}))$.

\subsection{$a$--maximization in the general case}

As explained in Section 3, the $Y^{p,q}$ family is obtained from the
$Y^{p,p}\simeq\C^3 /\IZ_2\times \IZ_p$ model by a sequence of $(p - q)$ 
simple modifications. Following this construction, and applying
the same logic presented for $Y^{3,2}$, 
it is straightforward to obtain a parametrization of the R--charges in
the general case.
There are then $2p$ relations from imposing the vanishing of the beta
functions at each node, and $p+q$ relations from requiring that each term in the
superpotential (again, by each term, we mean each $SU(2)$ doublet)
has R--charge 2. These are $3p+q$ linear relations in all, for
$(p-q)+p+q+(p+q)=3 p + q$ {\it a priori} independent R--charges. However,
two of the relations are redundant, and we can therefore
parameterize the R--charges of all fields  in terms of two unknowns $x$ and $y$ as follows:
\begin{itemize}
\item The $(p-q)$ singlets $Z$ around the outer loop of the
quiver have R--charge $x$.
\item The $(p+q)$ diagonal singlets $Y$ have R--charge $y$.
\item The $p$ doublets $U$
around the outer loop have R--charge
$1-\frac{1}{2}(x+y)$.
\item The $q$ doublets $V$ around the outer loop
have R--charge $1+\frac{1}{2}(x-y)$.
\end{itemize}

As already noted, the fact that the maximization is performed over a two dimensional space implies that there 
are precisely two $U(1)$ symmetries with which the R--symmetry can mix. 
It now follows that
\be
\mathrm{tr} R(x,y) = 2 p + (p - q) (x - 1) + (p+q)(y-1)-  p (x+y) +  q (x-y)  = 0
\ee
where $\mathrm{tr} R$ is a fermionic trace and the first contribution of $2 p$ comes from the gauginos.
We thus have tr$R=0$. This fact is always true for a theory with a weakly coupled supergravity dual \cite{henningson} (this corresponds to having $c = a$). In \cite{Benv-Hanany} a general proof is given 
that shows that tr$R$ vanishes for any superconformal quiver.
We can now compute tr$R^3(x,y)$ which reads
\bea
\mathrm{tr}R^3(x,y) = 2p +(p-q)(x-1)^3+ (p+q)
(y-1)^3-\frac{p}{4}(x+y)^3  +\frac{q}{4} (x-y)^3~.
\eea
The maximum is found at
\bea
y_{max} & = & \frac{1}{3q^2} \left[-4p^2+2pq+3q^2+(2p-q)\sqrt{4p^2-3q^2}\right]\nn\\
x_{max} & = & \frac{1}{3q^2} \left[-4p^2-2pq+3q^2+(2p+q)\sqrt{4p^2-3q^2}\right]~.
\eea
Notice immediately that these are precisely the same as the baryon
charges $R[B_1]$, $R[B_2]$ (\ref{baryons}) computed using the metrics.
Moreover, substituting into $a$ we also reproduce the correct
volume formula (\ref{volume}) via the AdS/CFT formula
\bea
a(Y^{p,q})=\frac{\pi^3}{4\cdot \vol(Y^{p,q})}~.\eea

\subsection{Continuous global symmetries}

We are now in a position to summarise the results obtained so far and conclude our comparison 
between geometric and field theoretic results.

In section \ref{sec:quivers} we constructed the quivers and wrote down the explicit superpotential. This superpotential is toric and satisfies a global $SU(2)$ symmetry. All the fields are in the spin--$0$ representation of $SU(2)$ or in the spin--$1/2$ representation. We note that applying successive Seiberg Dualities one expects to find higher dimensional representations.

In this section, solving the linear beta--function constraints, we showed that there are precisely two global $U(1)$ 
symmetries. For any $Y^{p,q}$ quiver the rank of the (antisymmetric) quiver matrix is $2 p - 2$ and the number of nodes 
is $2 p$. This implies that there is precisely one $U(1)$ \emph{baryonic} symmetry. The reason is that this baryonic
symmetry turns out to be equal to a particular combination of the $U(1)$
factors of the original $U(N)^{2 p}$ gauge groups. There are two symmetries that can be constructed by linear combinations of these $2 p$ $U(1)$ factors: one is completely decoupled and the other one is the baryonic symmetry. One way of checking that this is a baryonic symmetry is by computing the cubic 't Hooft anomaly, that has to vanish. The reason is that in the AdS dual description this global symmetry becomes a gauge symmetry, and the gauge field is given by Kaluza Klein reduction of the RR four--form of Type IIB superstrings on a $3$--cycle of the transverse 5--dimensional space. The cubic 't Hooft anomalies correspond to a Chern--Simons term in AdS that does not exist for gauge fields coming from RR four--forms. 

As a result of the previous discussion, we see from the gauge theory that any $Y^{p,q}$ quiver has to have precisely one $U(1)$ \emph{flavor} symmetry. For this symmetry one does not expect the cubic anomalies to vanish. This symmetry is related to the $U(1)$ part of the isometries of the transverse $5$--dimensional manifold. We summarise the final charges in table \ref{charges}.

\begin{table}[!h]
\begin{center}
$$\begin{array}{|c|c|c|c|c|}  \hline
\mathrm{Field}&\mathrm{number}&R -\mathrm{charge}& U(1)_B &  U(1)_F \\ \hline\hline
    Y      & p + q & (- 4 p^2 + 3 q^2 + 2 p q + (2 p - q)\sqrt{4 p^2 - 3 q^2})/3 q^2 & p - q &- 1 \\ \hline
    Z      & p - q & (- 4 p^2 + 3 q^2 - 2 p q + (2 p + q)\sqrt{4 p^2 - 3 q^2})/3 q^2 & p + q &+ 1 \\ \hline
U^{\alpha} &  p    & (2 p (2 p -  \sqrt{4 p^2 - 3 q^2}))/3 q^2                       & - p   & 0  \\ \hline
V^{\alpha} &  q    & ( 3 q - 2 p + \sqrt{4 p^2 - 3 q^2})/3 q                         &   q   &+ 1 \\ \hline
\end{array}$$
\caption{Charge assignments for the four different types of fields present in the general quiver diagram for $Y^{p,q}$.}
\label{charges}
\end{center}
\end{table}
%


%

From table \ref{charges} it is straightforward to compute
\be
\mathrm{tr} U(1)_B = \mathrm{tr} U(1)_F = 0~.
\ee
These linear anomalies have to vanish, since we have resolved the mixing with the R--symmetry.
A simple computation shows that also the cubic 't Hooft anomalies for $\mathrm{tr} U(1)^3_B$ and $\mathrm{tr} U(1)^3_F$ vanish, for instance
\be
\mathrm{tr} U(1)_B^3 = (p + q) (p - q)^3 + (p - q) (p + q)^3 + 2 p (- p)^3 + 2 q ( q )^3 = 0~.
\ee
The mixed 't Hooft anomalies $\mathrm{tr} U(1)_F^2 U(1)_B$ and $\mathrm{tr} U(1)_B^2 U(1)_F$ are instead non--zero. We note that the relation $\mathrm{tr} U(1)^3_F = 0$ does not have a direct physical meaning, since one can always redefine the flavor symmetry, mixing it with the baryonic symmetry. In this case, $\mathrm{tr} U(1)^3_F$ does not necessarily have to vanish. 

It is worth explaining the reason for the claim that $U(1)_B$, as given in table \ref{charges}, corresponds to a \emph{baryonic} symmetry, since the cubic anomalies $U(1)_B$ and $U(1)_F$ show a similar behaviour. One possible explanation is that $U(1)_B$ can be directly constructed as a linear combination of the $2 p$ decoupled gauge $U(1)$s, while $U(1)_F$ cannot. In the next subsection we will give a different explanation of this fact.


\subsection{Some properties of the baryons}

In this subsection we give a simple analysis of the baryonic operators in the $Y^{p, q}$ quivers, along the lines of \cite{Berenstein:2002ke,Intriligator:2003wr,Herzog:2003wt,Herzog:2003dj}.

Since all the $2 p$ gauge groups have the same rank, $N$, it is possible to construct simple dibaryonic operators with one type of bifundamental field $A_{\alpha}^{\;\, \beta}$:
\be\label{baryons1}
\mB [ A ] = \varepsilon^{\alpha_1 \ldots \alpha_N} \; A_{\alpha_1}^{\;\, \beta_1} \, 
            \ldots \, A_{\alpha_N}^{\;\, \beta_N} \; \varepsilon_{\beta_1 \ldots \beta_N}~.
\ee
In the $Y^{p, q}$ quivers there are four classes of bifundamental fields, so there are four classes of dibaryonic operators: $\mB[Y]$, $\mB[Z]$, $\mB[U]$ and $\mB[V]$. Since the fields $U^{\alpha}$ and $V^{\alpha}$ transform in the $2$--dimensional representation of the global $SU(2)$, the corresponding baryonic operators
transform in the $(N+1)$--dimensional representation, as explained in \cite{Berenstein:2002ke}. This fact tells us immediately that the corresponding D3--brane (wrapping a supersymmetric $3$--cycle inside the $Y^{p, q}$ manifold) can be freely moved on the round $S^2$ parametrized by the coordinates $\theta$ and $\phi$. The corresponding $3$--cycle is thus part of a family of supersymmetric cycles parametrized by an $S^2$.

The operators like (\ref{baryons1}) are chiral, so their scaling dimension is precisely the scaling dimension of the bifundamental $A$, multiplied by $N$. These scaling dimensions correspond holographically to the volumes of the corresponding $3$--cycles. Computations of the volumes give precisely the values listed in table \ref{charges}.

From the quiver it is also possible to derive some information about the topology of the $Y^{p, q}$ manifolds and of the supersymmetric $3$--cycles. A more complete treatment would require the algebraic computation of the moduli space of vacua of the gauge theory, which should reproduce the quotient of $\IC^4$ determined in \cite{DJ} and described in section \ref{sec:quivers}.

Taking the product of two different consecutive dibaryons it is possible to get rid of the two $\varepsilon$--symbols corresponding to the same gauge group \cite{Berenstein:2002ke}. For instance (for $q < p$) we can compose $U$--dibaryons with $Z$--dibaryons:
\be\label{baryons2}
\mB [ U ] \mB [ Z ] \sim \varepsilon^{\alpha_1 \ldots \alpha_N} \; 
 U_{\alpha_1}^{\;\, \beta_1} \, \ldots \, U_{\alpha_N}^{\;\, \beta_N} 
 Z_{\beta_1}^{\;\, \gamma_1} \, \ldots \, Z_{\beta_N}^{\;\, \gamma_N} 
\; \varepsilon_{\gamma_1 \ldots \gamma_N} \sim \mB [ U Z ]~.
\ee
It is thus possible to associate (poly--)baryonic operators to connected paths in the quiver diagram. When the path closes all the $\varepsilon$--symbols disappear and the operator is not baryonic anymore.

We thus look at the closed oriented paths in the quiver diagram. From the quiver diagrams we can recognize four different types of simple loops:
\begin{itemize}
\item One type of loop has length $3$ and is made of one $Y$--field, one $U$--field and one $V$--field. 
\item One type of loop has length $4$ and is made of one $Z$--field, two $U$--fields and one $Y$--field. 
\item The third type of loop instead goes all the way around the quiver and has length $2 p$: it is made of $p$ $U$--fields, $q$ $V$--fields and $p - q$ $Z$--fields. 
\item The last type has length $2 p - q$ and is made of $p$ $Y$--fields and $p - q$ $U$--fields.
\end{itemize}

In order that the interpretation of a closed path of baryonic operators as 
a non--baryonic operator makes sense, it is necessary that the total baryonic
charge vanishes. If we substitute the charges of table \ref{charges} into the
four types of closed loops listed above, we find that the two ``short'' loops
have vanishing charge both for the $U(1)_B$ and the $U(1)_F$ (as has to be the 
case since they enter in the superpotential), but the charge of the two
 ``long'' loops is zero only for $U(1)_B$. This implies that precisely the 
 symmetry called $U(1)_B$ in table \ref{charges} is the baryonic symmetry.

The fact that a closed loop of baryons is equivalent to a non--baryonic 
operator has an interpretation in terms of the topology of the $3$--cycles 
wrapped by the corresponding D3--brane: the sum of the cycles associated to 
the dibaryons entering the loop has to be the topologically trivial $3$--cycle.
Denoting $\Sigma [ A ]$ the $3$--cycle associated to the dibaryons constructed with
 the bifundamental $A$, we thus have the following four relations for the 
 corresponding homology cycles of the $Y^{p, q}$ manifold:
\bea
\label{Vdef} & \Sigma [ Y ] + \Sigma [ U ] + \Sigma [ V ] = 0 \\
\label{baserel} & \Sigma [ Z ] + 2 \, \Sigma [ U ] + \Sigma [ Y ] = 0 \\
\label{fromGysinone} & p \, \Sigma [ U ] + q \, \Sigma [ V ] + (p - q) \, \Sigma [ Z ] = 0 \\
\label{fromGysintwo} & p \, \Sigma [ Y ] + (p - q) \, \Sigma [ U ] = 0~. 
\eea
Recall that using the results of \cite{DJ} one can see that  for the singlet
dibaryons
%
$\Sigma [ Z ] = \Sigma_2 = S^3/\IZ_{p - q}$ and 
$\Sigma [ Y ] = \Sigma_1 = S^3/\IZ_{p + q}$. Moreover in \cite{DJ} it is shown that 
the representative cycles,  given by $\{y=y_i\}$ respectively, are supersymmetric,
meaning that the cones over these are complex divisors of the Calabi--Yau cones.

We will now show the existence of two new supersymmetric three--cycles,
corresponding precisely to the two remaining dibaryons, $\Sigma [ U ]$ and 
$\Sigma [ V ]$. 
First, let us verify that  we can pick a representative cycle of $\Sigma [U]$, 
which is supersymmetric 
and reproduces the correct volume/charge formula. This is again
straightforward using the results of \cite{DJ}. Consider 
the three--cycle obtained by setting $\{\theta,\phi\}$ to some constant value on the 
round $S^2$, and denote this by $\Sigma_3$. One easily computes
\bea
\frac{1}{2}J\wedge J |_{\{\theta,\phi\}=\mathrm{const.}} = \frac{1}{3} r^3 
\mathrm{d} r \wedge \mathrm{d} y \wedge \mathrm{d} \psi \wedge \mathrm{d} \alpha = 
\mathrm{vol}_{\{\theta,\phi\}=\mathrm{const.}} (\Sigma_3) \label{newvolume}
\eea
where $J$ is the K\"ahler two--form of the Calabi--Yau cone over 
$Y^{p,q}$ \cite{DJ}, and the induced volume form is computed using the metric 
(\ref{localmetric}). This shows that the cycle is supersymmetric. The topology is 
$S^3/\IZ_p$, as follows from the discussion  in \cite{DJ}, with the Chern number of the $U(1)$ fibration  over the $y-\psi$
two--sphere being $p$.
The volume is trivially computed by integrating (\ref{newvolume}) and indeed 
reproduces exactly $R[U]$. Finally, we may simply define the remaining cycle 
as the sum $\Sigma_4\equiv -\Sigma_1-\Sigma_3$. Thus from (\ref{Vdef}) 
we have $\Sigma[V]=\Sigma_4$.
Note that, correctly, 
vol$(\Sigma_4)\propto R[V]=R[U]+R[Z]$. Clearly the cycle is supersymmetric.

It is fairly straightforward to verify the topological relations (\ref{baserel})--
(\ref{fromGysintwo}) directly from the definitions of the cycles, thus 
providing a non--trivial check of the gauge theory calculation above. 
Let us first recall that $S_1$, $S_2$ are the two copies of $S^2$ in the 
base $B$ at $y=y_1$, $y=y_2$, respectively, and that $C_1$ is a copy 
of the fibre $S^2$ in $B$ at fixed $\theta$ and $\phi$. By definition, 
taking the $\alpha$ circle bundle over these submanifolds gives the 
3--cycles $\Sigma[Y]$, $-\Sigma[Z]$ and $-\Sigma[U]$, respectively\footnote{For 
further discussion of the topology the reader might consult \cite{DJ}.}. 
Recall now from \cite{paper2} that
\bea
S_1 - S_2 = 2C_1\eea%
holds as a homology relation in $B$. Thus taking the $\alpha$ circle bundle 
over this gives (\ref{baserel}).

Using (\ref{Vdef}) and (\ref{baserel}) one may now show that 
the left hand sides of (\ref{fromGysinone}) and (\ref{fromGysintwo}) are given by
\bea
p\Sigma[Y]+(p-q)\Sigma[U] & = & \frac{1}{2}\left((p+q)\Sigma[Y]-(p-q)\Sigma[Z]\right) \\ \nonumber
& = & -
\left(p\Sigma[U]+q\Sigma[V]+(p-q)\Sigma[Z]\right)~. \eea
Consider the quotient by $U(1)_{\alpha}$. As a homology relation 
in $B$ we have
\bea
\frac{1}{2}\left((p+q)\Sigma[Y]-(p-q)\Sigma[Z]\right)/U(1)_{\alpha} = \frac{1}{2}\left((p+q)S_1+(p-q)S_2\right) =  
pC_2 + qC_1\eea
where recall \cite{paper2} that by definition
\bea
S_1+S_2 = 2C_2\eea
and $C_1$ and $C_2$ are the canonical generators of the two two--cycles in $B\cong S^2 \times S^2$. 
Thus we must show that the $U(1)_{\alpha}$ bundle over $pC_2+qC_1$ is the trivial 3--cycle 
in $Y^{p,q}$.

To see this, we begin by noting that $\pi^*c_1$, the pull--back of the first 
Chern class of the $U(1)_{\alpha}$ bundle to the total space of $Y^{p,q}$, 
is trivial as a cohomology class. Here $\pi:Y^{p,q}\rightarrow B$ is the 
projection. This is a standard fact, and can be seen in a number of 
different ways. For example, the one--form $(\mathrm{d} \alpha+A)/2\pi\ell$ 
is globally defined on $Y^{p,q}$, where here $\mathrm{d}A/2\pi\ell$ is a 
representative for $c_1$ and $\alpha/\ell$ is a periodic coordinate on the 
$\alpha$ circle direction with period $2\pi$. The essential point is that
a gauge transformation on $\alpha$ is cancelled by the corresponding 
gauge transformation in $A$, thus giving a globally well--defined 
one--form on the total space -- the so--called global angular form. 
In particular, we note that the exterior derivative of this one--form, which represents
the pull--back of $c_1$, is exact.

To get to the desired homology relation, we simply apply 
Poincar\'e duality to the above. Since by definition $c_1 = p \sigma_1 
+ q\sigma_2$, where $\int_{C_i} \sigma_j = \delta_{ij}$ for each $i,j=1,2$, the
Poincar\'e dual to $c_1$ in $B$ is $pC_2+qC_1$. Following through 
\bea
H_2(B;\IZ)\cong H^2(B;\IZ)\stackrel{\pi^*}{\rightarrow} H^2(Y^{p,q};\IZ) 
\cong H_3(Y^{p,q};\IZ)\eea
then maps the two--cycle $pC_2+qC_1$ in $B$ to the 3--cycle in 
$Y^{p,q}$ which is simply the total space of the $\alpha$ circle bundle over this. 
As we've just explained, this image is zero.

\section{Conclusions}

The results of this paper change the {\it status quo} in AdS/CFT.
Until recently, the only explicitly known non--trivial Sasaki--Einstein 
metric in dimension five 
was $T^{1,1}$, whose 
 dual superconformal field theory -- a rather simple 
quiver gauge theory -- was given by Klebanov and Witten \cite{KW}. We now 
have an infinite number of explicit toric Sasaki--Einstein five--manifolds 
\cite{paper2}, their associated toric diagrams \cite{DJ}, and, from 
the results of this paper, we also have the whole infinite family 
of dual quiver gauge theories. This is remarkable.

We have applied the technique of 
$a$--maximization \cite{Intriligator:2003jj} to this infinite family of 
gauge theories to obtain the exact R--charges of the fields at the IR fixed 
point. 
As pointed out in \cite{Intriligator:2003jj}, since one is maximizing a cubic 
function with rational coefficients, the charges are generically 
quadratic irrational numbers, rather than rational numbers, 
and indeed this is typically true for 
the field theories presented here. There are also infinite numbers 
of theories where the R--charges are rational, namely the $Y^{p,q}$ quivers
with $4p^2-3q^2$ an integer square. The central 
charges of these theories, computed using field theory techniques, 
precisely match with the volumes computed using the explicit metrics 
found in \cite{paper2}. Furthermore, the R--charges of the gauge--invariant baryonic operators remarkably match the R--charges computed geometrically as volumes of supersymmetric cycles in the $Y^{p,q}$ geometries. 

In order to have a more complete picture it would be interesting to determine the moduli space of vacua of the $Y^{p,q}$ gauge theories, and reproduce the algebro--geometric results of \cite{DJ}.

Clearly this work opens the door to very interesting applications, and 
generalisations, in many different directions. First let us 
note that the construction of the Sasaki--Einstein metrics in 
\cite{paper2} was immediately generalised to all (odd) dimensions in 
\cite{paper3}. In particular, in dimension seven there are similar 
$(p,q)$ families of Sasaki--Einstein metrics which are based on 
any positive K\"ahler--Einstein metric in dimension four. These are classified, 
and consist of $\C P^2$, $S^2 \times S^2$ and the del Pezzo 
surfaces $dP_3,\ldots, dP_8$. These will therefore serve as supersymmetric M--theory 
backgrounds of the form $AdS_4\times Y_7$, which are expected  to be 
dual to $\mathcal{N}=2$ superconformal field theories arising on M2--branes that 
probe the corresponding Calabi--Yau four--fold singularities. 
When the K\"ahler--Einstein is toric, the Calabi--Yau singularities 
are again toric. It would be interesting to try to develop 
methods to analyse the gauge theory duals of these Sasaki--Einstein 
manifolds.

Let us also recall that the 
entire family of solutions explored in this paper has a dual description in M--theory, where it 
uplifts to supersymmetric $AdS_5\times M_6$ solutions, with $M_6$ a complex 
$S^2$ bundle over $T^2\times S^2$ \cite{paper1}. However, there are many more solutions
presented in \cite{paper1}, with $M_6$ replaced by more general manifolds. It will be very interesting
to investigate if, guided by our results, one could explicitly construct the dual
four--dimensional superconformal field theories for these also. If so, this could shed considerable
light on the corresponding M5--brane theory, it least in a conformal regime.

Another promising avenue of research is to understand what the 
geometric dual of $a$--maximization is. It is remarkable that 
such a simple field theory calculation reproduces not only the 
volumes of the metrics, but also the volumes of supersymmetric 
cycles. It is fair to say that very little is known about 
non--regular Sasakian--Einstein geometry ---
the metrics $Y^{p,q}$ were the first examples. However, the field 
theory results presented here suggest that there do exist general 
results. It will be very interesting to pursue this direction further.

Finally, there are clearly many avenues left to explore purely for the results presented here. For instance, it would be interesting to understand the precise structure of Seiberg Dualities, related to the construction of Klebanov--Strassler type geometries and to cascading RG flows. 

\section{Acknowledgements}

We would like to thank Yang--Hui He, Chris Herzog, Pavlos Kazakopoulos and Brian Wecht for useful discussions.
S. B. wishes to acknowledge the kind hospitality of CTP, where this work has been completed.
D. M. and J. F. S. would like to thank Matteo Bertolini and Francesco Bigazzi
for earlier collaboration on related material and e-mail correspondence.
J. F. S. is supported by NSF grants DMS--0244464, DMS--0074329 and DMS--9803347. 
The research of S. F. and A. H. was supported in part by the CTP and LNS of
MIT and the U.S. Department of Energy under cooperative research agreement $\#$ DE--FC02--
94ER40818, and by BSF, an American--Israeli Bi--National Science Foundation. A. H. is also
indebted to a DOE OJI Award.

\bibliographystyle{JHEP}

\end{document}